\documentclass[lettersize,journal]{IEEEtran}
\usepackage{amsmath,amsfonts}
\usepackage{algorithmic}
\usepackage[colorlinks, citecolor=green]{hyperref}
\usepackage{array}
\usepackage[caption=false,font=normalsize,labelfont=sf,textfont=sf]{subfig}
\usepackage{textcomp}
\usepackage{stfloats}
\usepackage{url}
\usepackage{verbatim}
\usepackage{graphicx}
\usepackage{cite}

\def\BibTeX{{\rm B\kern-.05em{\sc i\kern-.025em b}\kern-.08em
    T\kern-.1667em\lower.7ex\hbox{E}\kern-.125emX}}
\usepackage{balance}
\usepackage{booktabs}
\usepackage{xcolor} 
\usepackage{setspace}

\begin{document}
\title{
Beyond Reactive Assistance: PV-Care Using Low-Density EEG and AI to Provide Proactive, Context-Aware Help for MCI
} 
\author{Simon L Liu, Manish Kumar Krishne Gowda
 \thanks{Simon L Liu is with Shanghai High School International Division, Shanghai. 
200231, China (e-mail: simonleo.liu@gmail.com), Manish Kumar Krishne Gowda was with Elmore Family School of Electrical and Computer Engineering, Purdue University, West Lafayette, IN 47907 USA. He is now with Apple Inc., Cupertino, CA 95014 USA (e-mail: mkrishne@purdue.edu).}
    \vspace{-15pt}    

}

\markboth{Submitted to IEEE Transactions on Human-Machine Systems}%
{How to Use the IEEEtran \LaTeX \ Templates}

\maketitle    

\begin{abstract}
The growing elderly population gives rise to an urgent need for intelligent support systems, particularly for individuals with Mild Cognitive Impairment (MCI). This paper presents PV-Care, a proactive AI-driven assistance scheme that integrates wearable electroencephalogram (EEG) sensing with visual environmental perception to provide real-time, context-aware voice assistance for MCI users. Unlike traditional assistant systems that passively wait for user commands, PV-Care actively initiates helpful interactions based on the user's detected brain \textcolor{black}{states, including \textit{Learning}, \textit{Memory Recall}, and \textit{Resting}}, using a novel deep neural architecture named Spatial and Frequency Refinement Network (SFR-Net). By combining EEG-based cognitive-state recognition with AI-based visual analysis, PV-Care generates structured ``4$W$-UT'' prompts to guide the output of large language models (LLMs). Simulation results and user studies validate the high accuracy of the proposed SFR-Net and the effectiveness of PV-Care's context-aware assistance. \textcolor{black}{These results indicate that PV-Care is a feasible and promising solution for MCI caring.}
\end{abstract}

    \vspace{-10pt}
\section{Introduction} \label{sec: intro}
\IEEEPARstart {T}{he} growing elderly population necessitates innovative approaches to assist senior individuals, especially those experiencing mild cognitive impairment (MCI). MCI patients often encounter problems such as recognizing people or  \textcolor{black}{forgetting how to get back home}. Thus, caring for those with MCI always demands personal companions, which requires considerable effort in terms of time, money, and patience \cite{frech2024}.

Human-machine collaboration techniques have garnered significant attention, especially with the advancements in Artificial Intelligence (AI) technologies. This collaboration promotes the idea that machines can serve as embodied AI assistants that meet human needs
and assist them with various tasks \cite{li2023}. AI companion robots, as explored by Clara et al. \cite{Companion_Robots_Loneliness_2023}, \textcolor{black}{showed that AI companion robots could mitigate loneliness} among elders, offering emotional support and improving their mental well-being. Additionally, Cantone et al. \cite{Enhancing_Elderly_Health_Monitoring_2023} proposed integrating AI with autonomous robots and sensors to achieve secure and independent living for elderly individuals, which has been particularly beneficial for MCI patients. Zhou et al. \cite{Assistant_Robot_Dementia_2022} demonstrated how an assistant robot could enhance the perceived communication quality for people with MCI. 

Despite these advancements, existing solutions for elder care typically rely on users voluntarily issuing commands for interaction with AI robots or services. Due to cognitive decline, MCI patients often cannot correctly command assistant systems when they need assistance. Therefore, it is crucial to monitor the needs of MCI patients and proactively provide suitable assistance \cite{liu2025personalized}.

\textcolor{black}{To provide proactive assistance, it is \textcolor{black}{important} to sense the user's brain state and analyze when supportive services are needed. Electroencephalography (EEG) signals have been widely adopted to analyze users' brain activities. According to Schumacher et al. \cite{schumacher2020quantitative}, EEG characteristics may differ between healthy individuals and MCI patients, but both groups still show distinguishable EEG patterns under different cognitive states, such as attention-related and memory-related states. Therefore, although the specific EEG features of healthy individuals and MCI patients may not be identical, using EEG signals to recognize different brain states remains a feasible approach. With appropriate data collection and recognition methods, EEG-based brain-state analysis can be further integrated into MCI assistance systems for proactive service.}

Various AI methods have been \textcolor{black}{applied to analyze} EEG data. For example, Yue et al. \cite{yue2024tf} proposed a temporal-frequency hierarchical transformer network to capture key information from EEG signals. Yao et al. \cite{yao2024emotion} introduced TCNN, which leverages positional encoding and multi-head attention to extract channel information. Wang et al. \cite{wang2022towards} utilized a CNN-LSTM architecture to progressively extract temporal-spatial features through temporal and spatial convolutions. \textcolor{black}{Besides these AI-based methods, Bhalerao et al. also proposed sparse spectrum-based and swarm sparse decomposition methods for EEG signal analysis \cite{Bhalerao2022SSDM}. These methods have been applied to various EEG-related applications, such as cognitive visual object classification \cite{Bhalerao2024MSSDM}, and multi-class motor imagery-based EEG-BCI classification \cite{Bhalerao2025ESSDM}.}

However, these EEG analysis methods typically rely on EEG headsets with 32 or more channels. A major practical challenge is the inconvenience of wearing such complex equipment. In practical applications, convenient wearable EEG devices can often only sense a few channels, such as 4 or 6. For an MCI assistance system used in daily life, convenient low-density EEG recognition is especially important. Therefore, to enable EEG-based MCI assistant systems under these constraints, more advanced solutions are demanded for robust recognition of brain states using \textcolor{black}{low-density} EEG devices.

\begin{figure*}[!t]
    \centering
   \includegraphics[width=0.9\linewidth]{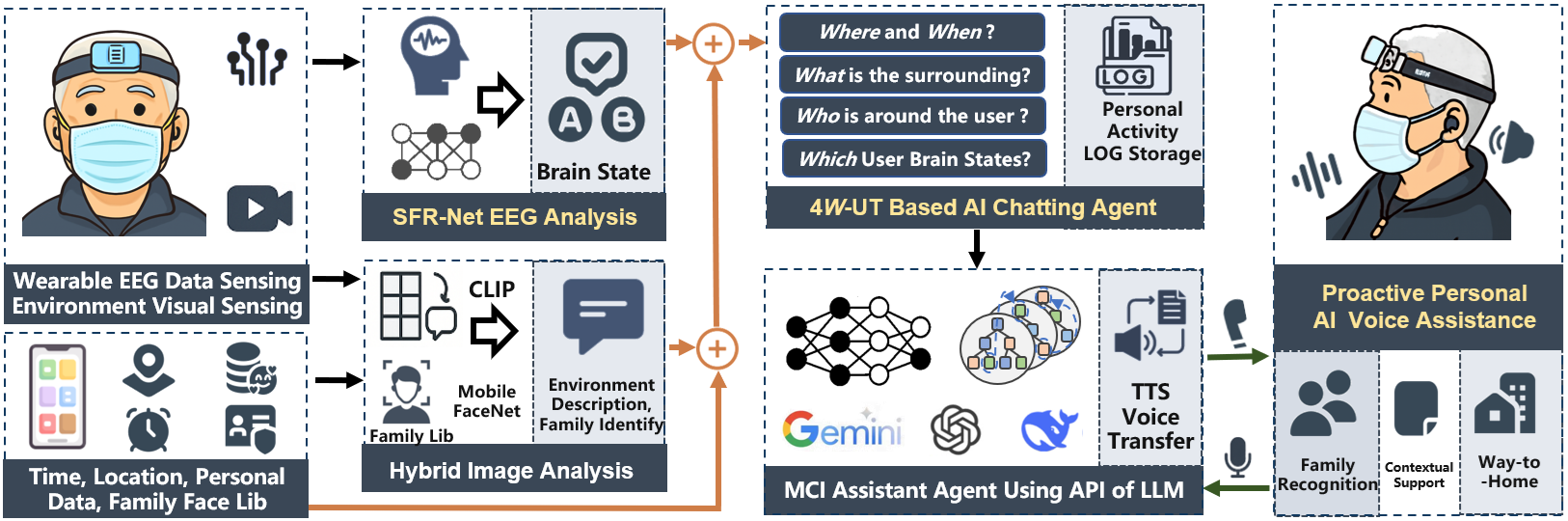}%
    \caption{The Proposed Proactive MCI Care Scheme: PV-Care}
    \label{fig:scheme}
    \vspace{-10pt}
\end{figure*}

\textcolor{black}{Large language models (LLMs) have shown strong capabilities in natural language understanding and generation, making them suitable for providing voice-based assistance to elderly users with cognitive impairment. Besides, recent advances in visual perception models and compact sensing hardware make it possible to analyze the user's surrounding environment in real time. Therefore, the integration of EEG-based cognitive-state analysis, visual environmental perception, and LLM-based dialogue provides a promising technical basis for building intelligent assistance systems that can understand both the user's brain activities and surrounding conditions.}

\textcolor{black}{However, existing assistance solutions still lack a unified framework that effectively integrates these techniques for proactive service. This paper proposes PV-Care, which uses EEG-based brain-state recognition as the trigger for proactive assistance and further combines environmental perception with an LLM-based agent to provide context-aware voice support for individuals with MCI. The main contributions of this work are summarized as follows:}

\begin{itemize}
\item We propose PV-Care, a proactive assistance scheme designed to support the daily lives of individuals with MCI. The scheme integrates EEG signal analysis, visual sensing, and an AI-based conversational agent that \textcolor{black}{simulates a familiar family member’s voice}. Specifically, PV-Care proactively delivers assistance based on the user's detected brain activity.

\item \textcolor{black}{We introduce an EEG-triggered LLM-based assistance mechanism for MCI users. SFR-Net is designed for low-density wearable EEG and recognizes the user's \textit{Resting}, \textit{Learning}, and \textit{Memory Recall} states. The recognized EEG state is used as the user-state (``UT'') component and integrated with the ``4$W$'' contextual information, including \textit{Where}, \textit{When}, \textit{Who}, and \textit{What}. The formed ``4$W$-UT'' prompt then triggers personalized assistance.}

\end{itemize} 

\vspace{-5pt}

\section{Our Proposal} \label{sec: method}

\subsection{Overall Architecture of the Proposed Proactive MCI Care Scheme }

Figure \ref{fig:scheme} illustrates the overall scheme of PV-Care, which including several key modules. The Environmental Visual Sensing module captures real-time images of the user's surroundings and employs image understanding methods such as \cite{openai_vision_api} to generate descriptive textual representations of the environment. \textcolor{black}{Additionally, when a person appears in the captured environmental images, a lightweight face recognition model, Mobile FaceNet \cite{FaceRecogAndroid}, is employed to identify user's familiar persons.}

\textcolor{black}{Meanwhile, the wearable EEG sensing module collects EEG data for SFR-Net to recognize the user's brain state, including \textit{Learning}, \textit{Resting}, and \textit{Memory Recall}. These three states are selected because they correspond to common and practically important conditions in daily life. The \textit{Resting} state indicates that the user may not need proactive assistance, which helps the system avoid unnecessary disturbance. The \textit{Learning} state is related to situations in which the user encounters and understands new information, while the \textit{Memory Recall} state is associated with scenarios in which the user attempts to remember important information or past activities. Considering the complexity of EEG signals and the limited amount of information provided by portable low-density EEG devices, this work focuses on accurately distinguishing these three representative states for practical proactive assistance.}

Textual descriptions from the visual module and cognitive state information analyzed from EEG data are integrated with additional contextual details, including time, location, and personal activity logs. These combined descriptions are then provided to a chatting agent powered by LLMs, such as ChatGPT, enabling contextually relevant and personalized interactions. By organizing the different functional modules in this way, PV-Care can proactively provide timely and precise assistance to MCI users based on their identified brain states.

For user convenience and comfort, EEG and visual sensing capabilities are integrated into a single wearable device paired with a smartphone. The smartphone’s earphones facilitate voice-based interactions between the user and the AI assistant. Using Text-to-Speech (TTS) technology, the assistant converts conversational texts to voice responses and delivers them via earphones. Furthermore, since patients with MCI often place greater trust in familiar family members, voice style transfer techniques \cite{voicetran} can be applied to simulate their voices, thereby enhancing user emotional engagement and trust in PV-Care. The detailed methods and implementations of each module are presented in the following sections.

    \vspace{-10pt}    

\subsection{Visual Sensing Module and Image-to-Environment Description}

\subsubsection{Visual to Textual Environment Description}
\begin{figure}[!t]
    \centering
    \includegraphics[width=0.85\linewidth]{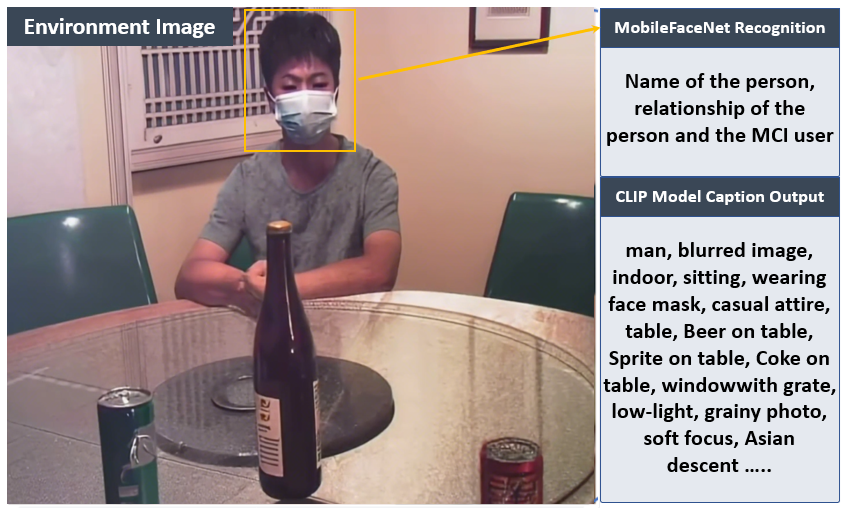}%
    \caption{Example of a sensed image and the corresponding environmental description generated by the cloud AI model.}
    \label{fig:yolo-test}
    \vspace{-10pt}

\end{figure}

In the proposed PV-Care scheme, the Visual Sensing Module captures real-time environmental images in front of the user, and the images are sent to the user’s smartphone and further forwarded for cloud-based image AI analysis. Recent advances in AI models \textcolor{black}{have demonstrated} remarkable performance in interpreting image content and converting it into textual descriptions. In our scheme, OpenAI’s GPT-4 Vision model is employed to analyze the environmental images, and the captured images are uploaded using the method described in \cite{openai_vision_api}, which processes the images and generates descriptive captions. For example, as illustrated in Figure \ref{fig:yolo-test}, a captured image is converted into the caption:
``man, blurred image, indoor, sitting Sprite on table, Coke on table...''

\subsubsection{Smartphone-Based Face Recognition}
As shown in Figure \ref{fig:yolo-test}, when people appear in the environmental images, PV-Care should identify them. Typically, an MCI patient only needs to recognize a few dozen familiar individuals. To achieve this, PV-Care employs a smartphone-optimized lightweight face recognition model \cite{FaceRecogAndroid}. Users can configure a personal face database, enabling the smartphone to accurately and efficiently identify them. This local processing ensures both low latency and strong privacy protection.

Overall, the Visual Sensing Module captures the user’s surrounding environment. General objects and scenes \textcolor{black}{are analyzed} by the cloud-based model to generate descriptive captions, while the smartphone-based face recognition model accurately identifies familiar individuals in the scene.

\vspace{-10pt}

\subsection{Spatial and Frequency Refinement Network (SFR-Net) for EEG Analysis}

\textcolor{black}{PV-Care is proposed to provide proactive services based on the user’s brain activity.} For instance, when a user attempts to recall information, PV-Care can offer appropriate suggestions without the user's explicit command. Therefore, accurate analysis of brain activity is essential for enabling proactive assistance. In our framework, we use EEG signals to analyze three brain activity states: \textcolor{black}{\textit{Learning}, \textit{Resting}, and \textit{Memory Recall}.}

\subsubsection{EEG Signal Acquisition and Experimental Dataset}

EEG signal processing is inherently affected by both the sensing device and individual variability. To develop and validate an EEG recognition model suitable for low-density EEG devices, we designed a dedicated EEG data acquisition experiment using both low-density and high-density EEG devices. Specifically, EEG signals were recorded simultaneously using a portable Muse EEG headband \cite{muse}, which provides four channels (AF7, AF8, TP9, and TP10), and a standard 64-channel Neuracle EEG system \cite{Neuracle}. \textcolor{black}{The 64-channel EEG recordings were used to support data validation, while the 4-channel wearable EEG signals were used to train and evaluate the proposed SFR-Net model. To ensure temporal synchronization between the two devices, both EEG systems were sampled at 256 Hz, and their signals were aligned during data organization. Details of the EEG data acquisition experiment, including the EEG recording protocol, the visual-stimulus design used during data collection are presented in Section I of the Supplemental Material.}

\begin{figure*}[t!]
\centering
\includegraphics[width=0.9\linewidth]{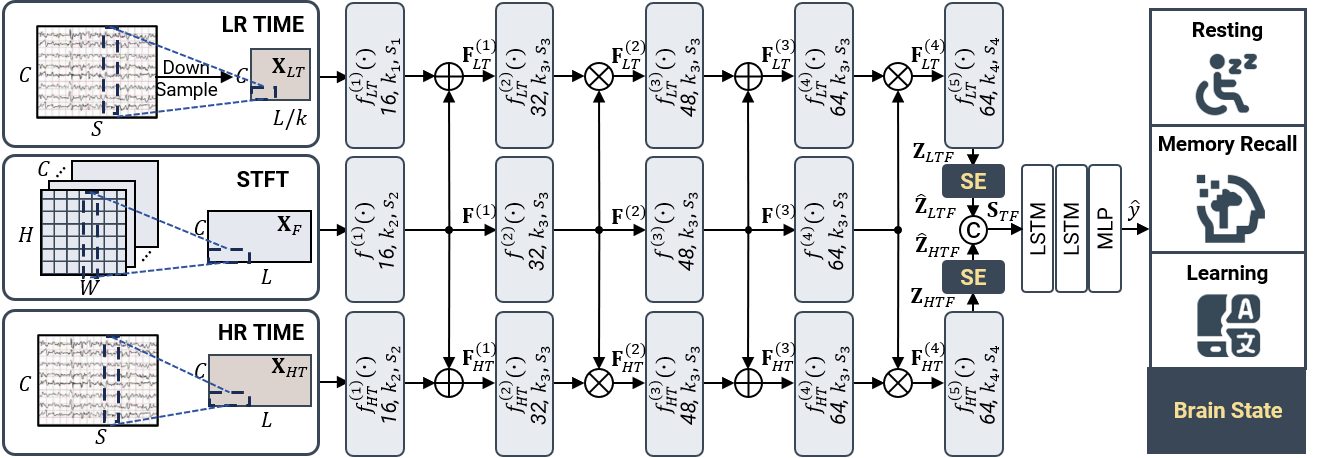}
\caption{Overall architecture of the SFR-Net for EEG classification.}
\label{fig:architecture}
    \vspace{-10pt}

\end{figure*}

\textcolor{black}{After completing the EEG data collection, we first verified the validity of the recorded data using the simultaneously collected 64-channel EEG signals. Specifically, following the classical EEG analysis methods in \cite{Friedman2004} and \cite{Neuner2014}, we examined whether the high-density EEG signals exhibited distinguishable patterns among the three states: \textit{Resting}, \textit{Learning}, and \textit{Memory Recall}. This analysis was used only as a data-validity check to confirm that the participants correctly followed the experimental protocol and that the recorded EEG segments contained state-related information. Once a segment was validated, the time-aligned 4-channel Muse EEG data collected during the same period were retained and used for training and evaluation of the proposed SFR-Net model.}


\subsubsection{Spatial and Frequency Refinement Network for EEG Analysis}

After validating the EEG signal dataset, we proposed a deep learning architecture named Spatial and Frequency Refinement Network (SFR-Net). This model analyzes 4-channel EEG signals and classifies three brain activity states—\textit{Resting}, \textit{Learning}, and \textit{Memory Recall}. The SFR-Net operates on 4-second EEG signal segments, with a preprocessing pipeline consisting of baseline correction, bandpass filtering (0.5–60~Hz), and artifact removal.

As shown in Figure~\ref{fig:architecture}, our SFR-Net model integrates multi-resolution temporal EEG signals with frequency-domain features through a multi-branch neural architecture, \textcolor{black}{which enables robust and hierarchical fusion} for enhanced brain state recognition. Let $C$ be the number of EEG channels (electrodes) and $L$ the number of time samples in a 4\,s segment sampled at $f_s=256$\,Hz, so $L=1024$. The proposed SFR-Net processes three complementary EEG representations as inputs:

\begin{itemize}
  \item High-resolution temporal signal \(\mathbf{X}_{{HT}} \in \mathbb{R}^{C \times L}\) (sampled at 256~Hz to capture fine-grained temporal dynamics).
  
    \item Frequency-domain features \(\mathbf{X}_{{F}} \in \mathbb{R}^{C \times L}\) are derived from the high-resolution temporal signal \(\mathbf{X}_{{HT}}\) via the Short-Time Fourier Transform (STFT). The STFT is applied to each channel of the input signal \(x(n)\) with a sampling rate of 256 Hz, employing a Hann window of length \(N = 256\) samples (corresponding to 1 second) and an overlap of 147 samples, resulting in a hop length of 109 samples. This yields 8 time frames for a 4-second EEG segment of length \(L = 1024\).
    
    The STFT is computed as:
    \begin{equation}
    \mathbf{X}(m, \alpha) = \sum_{n=0}^{N-1} x(n + mH) w(n) e^{-j 2 \pi \alpha n / N},
    \end{equation}
    where \(w(n)\) is the Hann window function, \(m\) is the frame index, \(H = 109\) is the hop length, and \(\alpha = 0, 1, \dots, N/2\) indexes the frequency bins (yielding 129 bins from 0 to 128 Hz).
    
    The magnitude spectrogram is computed, the DC component (0 Hz) is discarded, retaining 128 frequency bins. The resulting \(128 \times 8\) matrix per channel is normalized, transposed and flattened into a 1024-point vector to form \(\mathbf{X}_{{F}}\) across \(C\) channels.
  
  \item Low-resolution temporal signal \(\mathbf{X}_{{LT}} \in \mathbb{R}^{C \times L/k}\) (with \(k=4\)), \textcolor{black}{obtained} by downsampling the 256~Hz original signal to 64~Hz via average pooling to emphasize slower, global temporal trends. Here, \(C\) denotes the number of EEG channels, and \(L\) denotes the time steps in the high-resolution (HR) signal and the frequency bins.
\end{itemize} 

This multi-input design is motivated by the fusion of different brain activity patterns corresponding to EEG signal features: HR signals excel at detecting rapid neural transients (e.g., event-related potentials during learning), LR signals reduce noise and focus on sustained activities (e.g., resting baselines), and frequency features highlight rhythmic signatures (e.g., the $\alpha$ band of EEG is typically suppressed during memory recall).

To extract core temporal and spectral patterns and promote initial cross-modality alignment, the outputs are fused additively with the frequency branch \textcolor{black}{serving as an} anchor to infuse spectral context into temporal branches early on. This additive fusion is inspired by residual connections, enhancing gradient flow while integrating frequency priors to mitigate temporal aliasing. The initial convolution outputs are denoted as:
\begin{align}
\mathbf{F}^{(1)}_{{HT}} &= f^{(1)}_{{HT}}(\mathbf{X}_{{HT}}; \theta^{(1)}_{{HT}}) + \mathbf{F}^{(1)}, \\
\mathbf{F}^{(1)}_{{LT}} &= f^{(1)}_{{LT}}(\mathbf{X}_{{LT}}; \theta^{(1)}_{{LT}}) + \mathbf{F}^{(1)}, \\
\mathbf{F}^{(1)} &= f^{(1)}(\mathbf{X}_{{F}}; \theta^{(1)}),
\end{align}
where \(f^{(i)}_{*}(\cdot)\) represents the \(i\)-th convolutional block (Conv(\(\cdot\))) with learnable parameters \(\theta^{(i)}_{*}\) for the respective branch, and the subscript * denotes either HT, LT , or is omitted. Each branch begins with a convolutional block, comprising a convolutional layer, batch normalization (BN), ReLU activation, and dropout. Here, $f^{(1)}_{\mathrm{HT}}(\cdot)$ and $f^{(1)}(\cdot)$ are configured with 16 output channels, kernel size $k_{2}$, and stride $s_{2}$, whereas $f^{(1)}_{\mathrm{LT}}(\cdot)$ uses 16 output channels, kernel size $k_{1}$, and stride $s_{1}$. \(\mathbf{F}^{(1)}_{{HT}}, \mathbf{F}^{(1)}_{{LT}}, \mathbf{F}^{(1)} \in \mathbb{R}^{16 \times L_1}\) (with \(L_1\) adjusted post-convolution).

To enhance temporal-frequency interactions and capture mid-level features, a second convolutional block is applied, followed by multiplicative feature fusion. \textcolor{black}{The} gating mechanism, motivated by attention-like modulation, \textcolor{black}{enables} the frequency branch to selectively amplify salient temporal patterns, fostering adaptive integration across resolutions. The outputs are:
\begin{align}
\mathbf{F}^{(2)}_{{HT}} &= f^{(2)}_{{HT}}(\mathbf{F}^{(1)}_{{HT}}; \theta^{(2)}_{{HT}}) \otimes \mathbf{F}^{(2)}, \\
\mathbf{F}^{(2)}_{{LT}} &= f^{(2)}_{{LT}}(\mathbf{F}^{(1)}_{{LT}}; \theta^{(2)}_{{LT}}) \otimes \mathbf{F}^{(2)}, \\
\mathbf{F}^{(2)} &= f^{(2)}(\mathbf{F}^{(1)}; \theta^{(2)}),
\end{align}
where $f^{(2)}_{\mathrm{HT}}(\cdot)$, $f^{(2)}(\cdot)$, and $f^{(2)}_{\mathrm{LT}}(\cdot)$ are configured with 32 output channels, kernel size $k_{3}$, and stride $s_{3}$. The \(\otimes\) denotes element-wise multiplication calculation, and \(\mathbf{F}^{(2)}_{{HT}}, \mathbf{F}^{(2)}_{{LT}}, \mathbf{F}^{(2)} \in \mathbb{R}^{32 \times L_2}\).

Building on this, a third convolutional layer further refines hierarchical representations, with additive fusion to reinforce multi-resolution coherence. This stage is motivated by the progressive abstraction in neural hierarchies, where deeper layers integrate broader contexts. The outputs are:
\begin{align}
\mathbf{F}^{(3)}_{{HT}} &= f^{(3)}_{{HT}}(\mathbf{F}^{(2)}_{{HT}}; \theta^{(3)}_{{HT}}) + \mathbf{F}^{(3)}, \\
\mathbf{F}^{(3)}_{{LT}} &= f^{(3)}_{{LT}}(\mathbf{F}^{(2)}_{{LT}}; \theta^{(3)}_{{LT}}) + \mathbf{F}^{(3)}, \\
\mathbf{F}^{(3)} &= f^{(3)}(\mathbf{F}^{(2)}; \theta^{(3)}),
\end{align}
where $f^{(3)}_{\mathrm{HT}}(\cdot)$, $f^{(3)}(\cdot)$, and $f^{(3)}_{\mathrm{LT}}(\cdot)$ are configured with 48 output channels, kernel size $k_{3}$, and stride $s_{3}$. \(\mathbf{F}^{(3)}_{{HT}}, \mathbf{F}^{(3)}_{{LT}}, \mathbf{F}^{(3)} \in \mathbb{R}^{48 \times L_3}\).

Finally, a fourth convolutional block extracts high-level fused features, culminating in multiplicative fusion for fine-grained modulation. This multi-stage alternating fusion (additive-multiplicative) is \textcolor{black}{a key innovation}, enabling dynamic recalibration and preventing information loss in heterogeneous EEG modalities. The outputs are:
\begin{align}
\mathbf{F}^{(4)}_{{HT}} &= f^{(4)}_{{HT}}(\mathbf{F}^{(3)}_{{HT}}; \theta^{(4)}_{{HT}}) \otimes \mathbf{F}^{(4)}, \\
\mathbf{F}^{(4)}_{{LT}} &= f^{(4)}_{{LT}}(\mathbf{F}^{(3)}_{{LT}}; \theta^{(4)}_{{LT}}) \otimes \mathbf{F}^{(4)}, 
\end{align}
where $f^{(4)}_{\mathrm{HT}}(\cdot)$, $f^{(4)}(\cdot)$, and $f^{(4)}_{\mathrm{LT}}(\cdot)$ are configured with 64 output channels, kernel size $k_{3}$, and stride $s_{3}$. \(\mathbf{F}^{(4)}_{{HT}}, \mathbf{F}^{(4)}_{{LT}} \in \mathbb{R}^{64 \times L_4}\).

To achieve a deeper integration of the multi-resolution temporal and frequency features beyond the stage-wise fusions, we introduce a final fusion step by  applying a fifth convolutional block. The fused features are defined as:
\begin{align}
\mathbf{Z}_{{LTF}} &= f^{(5)}_{{LT}}(\mathbf{F}^{(4)}_{{LT}}; \theta^{(5)}_{{LT}}), \\
\mathbf{Z}_{{HTF}} &= f^{(5)}_{{HT}}(\mathbf{F}^{(4)}_{{HT}}; \theta^{(5)}_{{HT}}),
\end{align}
where $f^{(5)}_{\mathrm{HT}}(\cdot)$ and $f^{(5)}_{\mathrm{LT}}(\cdot)$ are configured with 64 output channels, kernel size $k_{4}$, and stride $s_{4}$, and $\mathbf{Z}_{{LTF}}, \mathbf{Z}_{{HTF}} \in \mathbb{R}^{64 \times L_5}$.

The fused branch outputs are then processed through a Squeeze-and-Excitation (SE) block to emphasize channel-wise dependencies, motivated by the need to prioritize informative EEG channels. The SE-enhanced features are denoted as:
\begin{align}
\hat{\mathbf{Z}}_{{LTF}} &= \mathrm{SE}(\mathbf{Z}_{{LTF}}), \\
\hat{\mathbf{Z}}_{{HTF}} &= \mathrm{SE}(\mathbf{Z}_{{HTF}}),
\end{align}
where \(\mathrm{SE}(\cdot)\) applies global average pooling followed by a two-layer MLP for excitation weights, and \(\hat{\mathbf{Z}}_{{LTF}}, \hat{\mathbf{Z}}_{{HTF}} \in \mathbb{R}^{64 \times L_5}\).

\begin{figure*}[!t]
    \centering
    \includegraphics[width=1\linewidth]{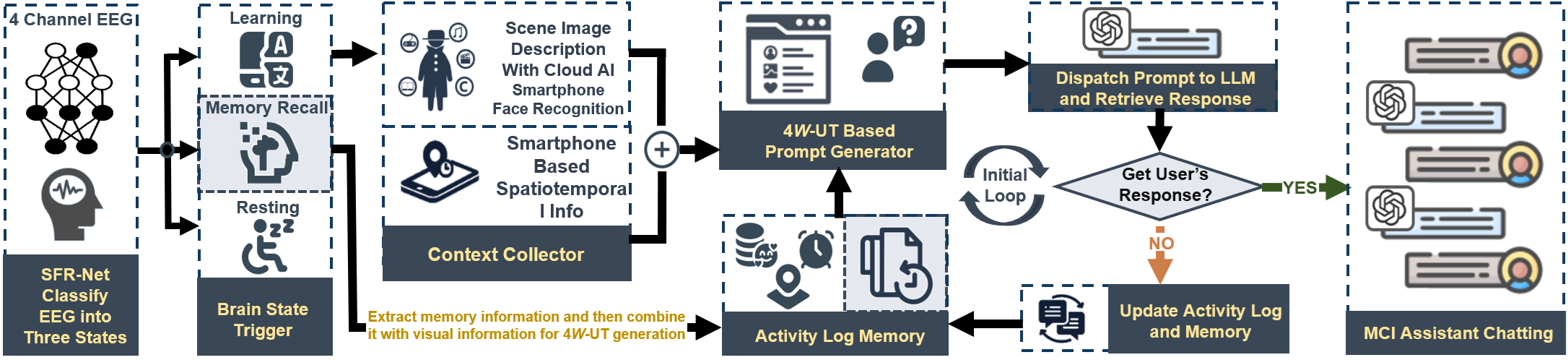}%
   \caption{\textcolor{black}{ Workflow of EEG-triggered ``4$W$-UT'' and personal-data-combined prompts for the LLM-based MCI chatting agent.} }
    \label{fig:flow}
    \vspace{-15pt}
    
\end{figure*}

To capture long-range temporal dependencies and bidirectional context in EEG sequences, we employ a bidirectional two-layer Long Short-Term Memory (LSTM) network. After max-pooling, the flattened features from \(\mathbf{Z}_{{LTF}}\) and \(\mathbf{Z}_{{HTF}}\) are processed by the LSTM to obtain the hidden representation:
\begin{align}
\mathbf{S}_{{TF}} &= \mathrm{Concat}(\hat{\mathbf{Z}}_{ {HTF}}, \hat{\mathbf{Z}}_{ {LTF}}), \\
\mathbf{H} &= \mathrm{LSTM}(\mathrm{Flatten}(\mathbf{S}_{ {TF}}); \phi),
\end{align}
where \(\mathrm{Concat}(\cdot)\) is channel-wise concatenation, \(\mathrm{Flatten}(\cdot)\) vectorizes the tensor, \(\mathrm{LSTM}(\cdot; \phi)\) is the two-layer bidirectional LSTM with parameters \(\phi\), and \(\mathbf{H} \in \mathbb{R}^{256}\).

Finally, a Multi-Layer Perceptron (MLP) with two fully connected layers (hidden size 256, ReLU activation, dropout 0.5) performs the classification, outputting probabilities for the three brain states. This MLP is motivated by its efficiency in non-linear decision boundaries for multi-class tasks. The prediction is:
\begin{align}
\hat{y} = \mathrm{Softmax}(\mathrm{MLP}(\mathbf{H}; \psi)),
\end{align}
where \(\mathrm{MLP}(\cdot; \psi)\) denotes the MLP with parameters \(\psi\), and \(\hat{y} \in \mathbb{R}^{3}\).

\textcolor{black}{Using the accurately labeled 4-channel EEG dataset established in the previous section, the proposed SFR-Net model is trained end-to-end with cross-entropy loss. 
The model adopts a multi-branch, multi-stage fusion strategy to capture spatial, spectral, and temporal dependencies in EEG signals, enabling robust classification of the three target cognitive states: \textit{Resting}, \textit{Learning}, and \textit{Memory Recall}. 
Within the PV-Care scheme, these recognized states are further used to guide proactive assistance. 
Specifically, the \textit{Learning} state triggers explanatory and environment-oriented assistance, while the \textit{Memory Recall} state triggers retrieval-oriented support, such as reminders of previous activities, persons, or routes.}

\subsection{``4$W$-UT'' and Personal Data Combined Prompt for LLM-Based MCI Chatting Agent}

Cloud-based AI systems, such as ChatGPT, have demonstrated remarkable capabilities in conversational tasks and problem-solving. However, a significant challenge in using LLMs for MCI assistance is the variability and potential hallucination of generated responses; therefore, well-designed prompts are required to guide and constrain the LLM's output.

\subsubsection{``4$W$-UT'' Prompt for Chatting Setup}

In the PV-Care scheme, we designed a chatting agent, the pipeline of which is shown in Figure~\ref{fig:flow}. Prompt-based control, such as \textit{``role: system, content: descriptions''}, is used in our agent to initialize the cloud LLM. Specifically, we define the ``4$W$-UT'' prompt, which integrates multi-source context to drive personalized conversations. The key information of ``4$W$-UT'' prompt are:

\begin{itemize}
    \item \textit{Where}: the user’s location, derived from the smartphone GPS (position format: ddmm,N/S, dddmm,E/W).
    \item \textit{When}: the current time, also from the smartphone (time format: year-month-day-time).
    \item \textit{Who}: the people present, recognized locally using the smartphone-based face recognition system.
    \item \textit{What}: the surrounding environment description, generated by the cloud-based visual AI model (e.g., ``man, table, bottle on table, can on table, etc.'' as in Figure~\ref{fig:yolo-test}).
    \item User-State: the user’s cognitive state and personal log.  
    In addition to \textit{Resting}, \textit{Learning}, or \textit{Memory Recall}, recent activity history and personalized information of the MCI patient are incorporated. 
    
\end{itemize}

\textcolor{black}{By combining this information, the agent can proactively initiate a conversation using the } ``4$W$-UT'' prompt: 
For example shown in Figure~\ref{fig:yolo-test},  \textcolor{black}{assuming that the person in the image is named Leonardo}. Based on the collected contextual information, a generated ``4$W$-UT'' prompt that is sent to the OpenAI API might be:
\begin{quote}
\textit{role: system, content: ``I am in a room with GPS ddmm,N/S, dddmm,E/W. Current time is year-month-day-time. Leonardo is here. In front of me: a man, indoors, sitting, wearing a face mask, casual attire, and a table with bottles and cans. The user is currently in a memory recall state.''}
\end{quote}

Additionally, our scheme incorporates the user’s recent weekly activity data and planned next-step actions. For example, if an MCI patient leaves home in the morning, they may forget to return. By maintaining an activity log and integrating it into the prompt, \textcolor{black}{PV-Care’s chatting agent can provide reminders} during \textit{Memory Recall} states, such as suggesting the user should return home and offering the navigation route.


\subsubsection{The Workflow of Our Assistant Chatting Agent}
Figure~\ref{fig:flow} presents the overall workflow of our chatting agent. \textcolor{black}{The input of the agent is the user's brain states which are classified by SFR-Net into three states, and the chatting pipeline is as follows:}

\begin{itemize}
    \item Brain State Trigger: 
    This module receives the brain activity state outputs from the SFR-Net, as described in the previous section. The \textit{Memory Recall} and \textit{Learning} states activate subsequent modules to proactively initiate user interactions, while the \textit{Resting} state does not trigger the assistance.

    \item Context Collection: 
    Multi-source context is collected, including environmental descriptions generated by the cloud model from the user’s surroundings, as well as the identification of potential people using the user’s smartphone-based face recognition. This completes the gathering of necessary environmental information.

    \item Activity Log Memory:
    This module logs the user’s recent activities, such as ``left home at 9 AM'',  which provides contextual information for subsequent interactions. This log is particularly useful when the user is in the \textit{Memory Recall} state, enabling the system to create relevant prompts based on past activities.

    \item ``4$W$-UT'' Based Prompt Generator: 
    If the user is in a \textit{Memory Recall} state, the activity log memory information is fused with environmental data to form the ``4$W$-UT'' prompt. If the user is not in a \textit{Memory Recall} state, the prompt is constructed solely from the environmental image recognition descriptions used for the \textit{Learning} state.

    \item Dispatch Prompt and Feedback Loop:
    \textcolor{black}{The constructed prompts are sent to the LLM as system-role instructions, i.e., \textit{``role: system''}, to initiate a proactive assistance dialogue. After each prompt is delivered, the system waits for 10 seconds to monitor whether the user provides feedback. If no response is received within this period, the system updates the contextual information and re-enters the prompt generation process to initiate another assistance attempt. This process is repeated for up to three attempts to avoid excessive disturbance to the user. If the user still provides no feedback after three attempts, the current dialogue attempt is terminated, and the system waits for the next EEG-triggered activation before providing further assistance. Once a meaningful user response is received, the LLM continues the interaction and provides context-aware support.}
\end{itemize}

\textcolor{black}{Using the workflow shown in Figure~\ref{fig:flow}, PV-Care can provide a highly personalized and context-aware conversational experience. Moreover, the textual responses generated by the LLM can be transformed into speech using TTS technology to provide voice-based assistance.}

\section{Simulation Results and Discussion} \label{sec: experiments}

In the experimental evaluation, we conducted functional simulations of the proposed proactive MCI assistance scheme. As illustrated in Figure~\ref{fig:scheme}, the PV-Care scheme integrates three core components: scene image analysis, EEG-based brain state analysis, and LLM-based chatting agent. For the image analysis, we directly utilized OpenAI’s API. Hence, our experiment primarily focused on the EEG-based brain state recognition and the proactive chatting simulations.

\vspace{-10pt}    

\subsection{Evaluation of SFR-Net for EEG-Based Brain State Classification}

To evaluate the effectiveness of the proposed SFR-Net model, we conducted experiments on the 30-subject EEG dataset described in Section II.C and Section I of the Supplemental Material.
In this initial feasibility study, EEG data were collected from healthy participants because the \textit{Resting}, \textit{Learning}, and \textit{Memory Recall} tasks can be performed more reliably, and the data acquisition process can be better controlled. 
\textcolor{black}{Although EEG characteristics of individuals with MCI may differ from those of healthy participants, prior studies also indicate that MCI-related EEG signals still contain distinguishable state-dependent patterns~\cite{schumacher2020quantitative}. 
Therefore, the purpose of this experiment is to verify that the three target cognitive states are separable and that the proposed SFR-Net is capable of learning discriminative representations for state classification. With appropriate training data, the same network architecture can be retrained or fine-tuned for MCI-specific EEG state recognition.}

\subsubsection{Experimental  Setup}

\textcolor{black}{The 4-channel EEG signals were preprocessed and segmented into 4-second windows sampled at 256 Hz, resulting in an input size of $1 \times 4 \times 1024$ for each sample. Adjacent segments were generated with an overlap ratio of 0.5. We adopted 5-fold cross-validation to improve the reliability of the evaluation and reduce the impact of data partitioning. The SFR-Net model was implemented in PyTorch and trained on an NVIDIA GeForce RTX 4090 GPU using the Adam optimizer with a batch size of 32 and a learning rate of 0.001. The model was trained for up to 300 epochs, with the dropout rate set to 0.5. The final performance on the test set is reported using average accuracy, precision, recall, and F1 score.}

\subsubsection{Comparison with State-of-the-Art Methods}

\textcolor{black}{We compared SFR-Net with nine representative EEG classification baselines, including EEGNet~\cite{Lawhern2018-ah}, Tsception~\cite{Ding2023-yl}, Conformer~\cite{Song2023-zu}, MSTCNN~\cite{Liu2023-tp}, CNNLSTM~\cite{Masuda2023-ck}, BMFCNet~\cite{10908708}, TS-SEFFNet~\cite{Li2021TSSEFFNet}, TF-HybridNet~\cite{Sui2021TFHybridNet}, and CE-stSENet~\cite{Li2020CEstSENet}. These baselines cover lightweight CNN-based models, multi-scale temporal--spatial networks, CNN--RNN hybrid architectures, transformer/attention-based methods, and temporal--frequency fusion models, providing a comprehensive comparison for low-density EEG classification.}

\begin{table}[!t]
\centering
\caption{\textcolor{black}{Performance Comparison of Different Methods}}
\label{tab:results}
\scriptsize{
\setlength{\tabcolsep}{2.8pt} 
\begin{tabular}{@{}lcccc@{}}
\toprule
\textbf{Method} & \textbf{Acc} & \textbf{Precision} & \textbf{Recall} & \textbf{F1 Score} \\
\midrule
EEGNet      & 0.669$\pm$0.004 & 0.675$\pm$0.002 & 0.667$\pm$0.008 & 0.668$\pm$0.005 \\
Tsception   & 0.720$\pm$0.018 & 0.751$\pm$0.014 & 0.715$\pm$0.022 & 0.713$\pm$0.024 \\
Conformer   & 0.726$\pm$0.011 & 0.738$\pm$0.012 & 0.728$\pm$0.010 & 0.728$\pm$0.011 \\
MSTCNN      & 0.754$\pm$0.014 & 0.754$\pm$0.015 & 0.768$\pm$0.017 & 0.754$\pm$0.014 \\
CNNLSTM    & 0.781$\pm$0.013 & 0.800$\pm$0.020 & 0.789$\pm$0.016 & 0.782$\pm$0.013 \\
BMFCNet     & 0.786$\pm$0.019 & 0.779$\pm$0.018 & 0.799$\pm$0.019 & 0.785$\pm$0.020 \\
TS-SEFFNet  & 0.763$\pm$0.005 & 0.794$\pm$0.005 & 0.756$\pm$0.008 & 0.764$\pm$0.006 \\
CE-stSENet  & 0.665$\pm$0.005 & 0.678$\pm$0.006 & 0.673$\pm$0.004 & 0.664$\pm$0.005 \\
TF-HybridNet& 0.736$\pm$0.015 & 0.782$\pm$0.003 & 0.733$\pm$0.021 & 0.739$\pm$0.014 \\
\textbf{SFR-Net}   & \textbf{0.804$\pm$0.016} & \textbf{0.812$\pm$0.020} & \textbf{0.813$\pm$0.011} & \textbf{0.804$\pm$0.017} \\
\bottomrule
\end{tabular}
}
    \vspace{-10pt}
\end{table}

\begin{figure}[!t]
    \centering
    \includegraphics[width=1\linewidth]{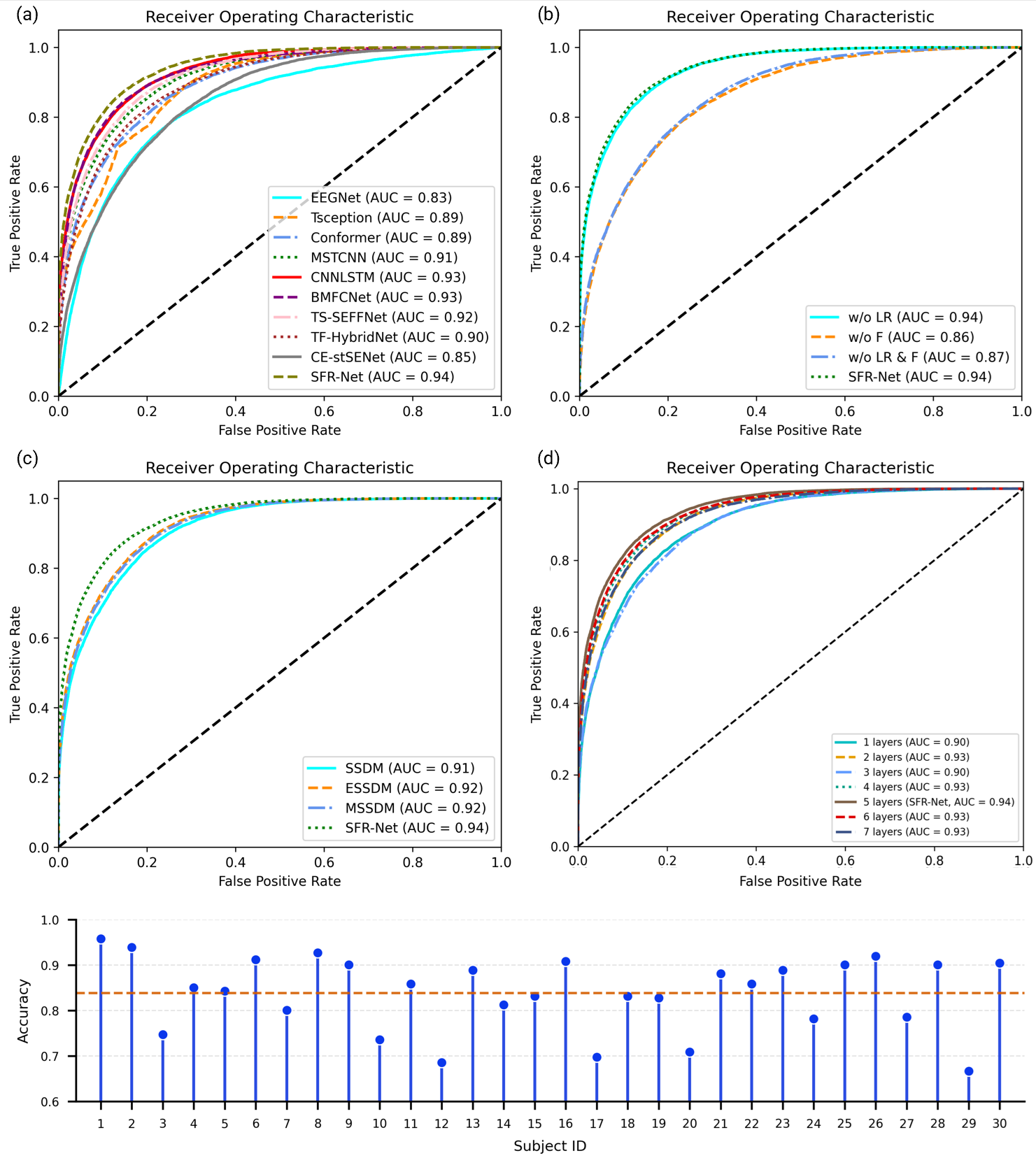}%
    \vspace{-5pt}
    \caption{\textcolor{black}{(a)  ROC of SFR-Net for three-state brain activity classification using 5-fold cross-validation, (b)  ROC analysis of ablation study for SFR-Net, (c) ROC curves of SSDM, ESSDM, MSSDM, and SFR-Net, (d) ROC curves of SFR-Net variants with different numbers of layers, (e) Subject-wise accuracy of SFR-Net across 30 subjects. The dashed line indicates the mean accuracy.} }
        \label{fig:roc_curves_compare}
 
    \end{figure}

As shown in Table~\ref{tab:results}, SFR-Net outperforms all baselines across all metrics, achieving the highest accuracy of 0.804$\pm$0.016. Notably, SFR-Net improves upon the closest competitor, BMFCNet, by approximately 1.8\% in accuracy, demonstrating the advantages of its hierarchical fusion of multi-resolution temporal and frequency-domain features. Lower-performing models rely primarily on single-resolution inputs, underscoring the limitations of ignoring complementary dynamics across time scales.

The superiority of SFR-Net is further evidenced by the ROC curves in Figure~\ref{fig:roc_curves_compare}(a), where it achieves the highest AUC of 0.94. This indicates robust separability, {with SFR-Net's curve closer to the top-left corner} than those of CNNLSTM and BMFCNet (both AUC 0.93). The marginal AUC gains highlight SFR-Net's enhanced ability to minimize false positives while maintaining high true positive rates, which is critical for real-world brain-computer interface applications where misclassification of cognitive states could degrade usability.

\begin{table}[t]
\centering
\caption{Ablation Study of SFR-Net Components}
\label{tab:ablation}
\scriptsize{
\setlength{\tabcolsep}{2pt} 
\begin{tabular}{@{}lcccc@{}}
\toprule
\textbf{Method} & \textbf{Acc} & \textbf{Prec.} & \textbf{Recall} & \textbf{F1} \\
\midrule
w/o LR      & 0.797$\pm$0.010 & 0.809$\pm$0.015 & 0.803$\pm$0.007 & 0.797$\pm$0.009 \\
w/o F       & 0.685$\pm$0.007 & 0.688$\pm$0.007 & 0.700$\pm$0.007 & 0.684$\pm$0.008 \\
w/o LR\&F  & 0.688$\pm$0.012 & 0.697$\pm$0.012 & 0.701$\pm$0.010 & 0.688$\pm$0.012 \\
\textbf{SFR-Net}   & \textbf{0.804$\pm$0.016} & \textbf{0.812$\pm$0.020} & \textbf{0.813$\pm$0.011} & \textbf{0.804$\pm$0.017} \\
\bottomrule
\end{tabular}
}
    \vspace{-10pt}
\end{table}

\subsubsection{Ablation Study}

To validate the contributions of key components in SFR-Net—specifically the low-resolution (LR) temporal branch and the frequency-domain (F) branch—we performed an ablation study by systematically removing these elements and retraining the model.

Results in Table~\ref{tab:ablation} reveal that both branches are essential. Removing the LR branch (w/o LR) results in a modest drop in accuracy to 0.797$\pm$0.010, suggesting that while high-resolution inputs capture fine details, the LR branch provides complementary stability for slower neural dynamics. More critically, ablating the F branch (w/o F) causes a substantial decline to 0.685$\pm$0.007 accuracy, emphasizing the importance of spectral features in encoding oscillatory patterns like theta-band activity during \textit{Memory Recall}. The combined removal (w/o LR\&F) yields performance comparable to w/o F (accuracy 0.688$\pm$0.012), confirming that frequency-domain integration is the dominant factor, though LR enhances it further in the full model.

The ROC analysis in Figure~\ref{fig:roc_curves_compare}(b) corroborates these findings, with the full SFR-Net achieving an AUC of 0.94. Interestingly, w/o LR maintains the same AUC (0.94), but the table metrics indicate reduced consistency, implying that LR aids in balanced multi-class performance rather than overall separability. In contrast, w/o F and w/o LR\&F drop to AUCs of 0.86 and 0.87, respectively, with curves deviating further from the ideal, highlighting how frequency fusion mitigates trade-offs in false positive rates. These results affirm the multi-branch fusion in SFR-Net as pivotal for superior EEG decoding.

\textcolor{black}{Since the frequency information plays a critical role, we further examined whether decomposition-based frequency representations could replace the original STFT input. Specifically, we incorporated three sparse decomposition-based methods into the ablation study: SSDM~\cite{Bhalerao2022SSDM}, ESSDM~\cite{Bhalerao2025ESSDM}, and MSSDM~\cite{Bhalerao2024MSSDM}. These methods represent sparse-spectrum-guided, enhanced, and multivariate swarm sparse decomposition strategies for extracting oscillatory or channel-aligned EEG components from non-stationary EEG signals.} \textcolor{black}{As shown in Table~\ref{tab:decomp_compare} and Figure~\ref{fig:roc_curves_compare}(c),  SFR-Net consistently outperforms these decomposition-based variants in both quantitative metrics and ROC evaluation. These results indicate that, although decomposition-based preprocessing methods are effective and relevant for non-stationary EEG analysis, the original STFT branch is more compatible with our multi-resolution temporal-frequency fusion architecture and low-density wearable EEG setting.}

\begin{table}[t]
\centering
\caption{\textcolor{black}{Performance Comparison with Decomposition-based Preprocessing Methods.}}
\label{tab:decomp_compare}
\scriptsize{
\setlength{\tabcolsep}{2pt}
\begin{tabular}{@{}lcccc@{}}
\toprule
\textbf{Method} & \textbf{Acc} & \textbf{Prec.} & \textbf{Recall} & \textbf{F1} \\
\midrule
SSDM    & 0.745$\pm$0.006 & 0.767$\pm$0.008 & 0.749$\pm$0.006 & 0.739$\pm$0.008 \\
ESSDM   & 0.764$\pm$0.009 & 0.779$\pm$0.008 & 0.772$\pm$0.007 & 0.762$\pm$0.010 \\
MSSDM   & 0.760$\pm$0.009 & 0.776$\pm$0.006 & 0.768$\pm$0.009 & 0.758$\pm$0.010 \\
\textbf{SFR-Net} & \textbf{0.804$\pm$0.016} & \textbf{0.812$\pm$0.020} & \textbf{0.813$\pm$0.011} & \textbf{0.804$\pm$0.017} \\
\bottomrule
\end{tabular}
}
\vspace{-5pt}
\end{table}


\begin{table}[t]
\centering
\caption{\textcolor{black}{Performance Comparison with Different Model Depths.}}
\label{tab:depth_compare}
\scriptsize{
\setlength{\tabcolsep}{1.8pt}
\begin{tabular}{@{}lcccccc@{}}
\toprule
\textbf{Layers} & \textbf{Acc.} & \textbf{Prec.} & \textbf{Recall} & \textbf{F1} & \textbf{Params} & \textbf{Time} \\
& & & & & \textbf{(M)} & \textbf{(ms)} \\
\midrule
1 & 0.739$\pm$0.007 & 0.753$\pm$0.004 & 0.748$\pm$0.007 & 0.739$\pm$0.008 & 2.158 & 0.560 \\
2 & 0.776$\pm$0.010 & 0.784$\pm$0.007 & 0.792$\pm$0.008 & 0.776$\pm$0.010 & 2.175 & 0.613 \\
3 & 0.731$\pm$0.008 & 0.745$\pm$0.008 & 0.741$\pm$0.008 & 0.730$\pm$0.007 & 2.207 & 0.717 \\
4 & 0.788$\pm$0.006 & 0.800$\pm$0.006 & 0.798$\pm$0.006 & 0.789$\pm$0.007 & 2.215 & 0.818 \\
\textbf{5} & \textbf{0.804$\pm$0.016} & 0.812$\pm$0.020 & \textbf{0.813$\pm$0.011} & \textbf{0.804$\pm$0.017} & 2.219 & 0.965 \\
6 & 0.800$\pm$0.004 & \textbf{0.813$\pm$0.007} & 0.808$\pm$0.005 & 0.800$\pm$0.004 & 2.268 & 1.026 \\
7 & 0.780$\pm$0.009 & 0.796$\pm$0.007 & 0.787$\pm$0.008 & 0.779$\pm$0.009 & 2.306 & 1.108 \\
\bottomrule
\end{tabular}
}
\end{table}

\begin{table}[t!]
\centering
\caption{Cognitive Assessment Summary for 60 Participants}
\label{tab:patient_scores}
\footnotesize
\setlength{\tabcolsep}{3pt}
\begin{tabular}{@{}lccccc@{}}
\toprule
\textbf{Metric} & \textbf{N} & \textbf{Age (Years)} & \textbf{Gender (Female)} & \textbf{MMSE} & \textbf{MoCA} \\
\midrule
\textbf{Value} & 60 & 68 $\pm$ 8 & 37 (61.7\%) & 26 $\pm$ 3 & 22 $\pm$ 6 \\
\bottomrule
\end{tabular}
    \vspace{-10pt}
\end{table}

\begin{table*}[!t]
\centering
\caption{Chat Prompt Evaluation Metrics for PV-Care, Including Subjective (M1--M4) and Objective (M5) Indicators.}
\label{tab:chat_eval_metrics}
\scriptsize{
\setlength{\tabcolsep}{1.5pt} 
\renewcommand{\arraystretch}{1.05} 
\begin{tabular}{p{0.04\linewidth}p{0.23\linewidth}p{0.45\linewidth}p{0.18\linewidth}p{0.06\linewidth}}
    \toprule
\textbf{No.} & \textbf{Metric Name} & \textbf{Description} & \textbf{Evaluation Method} & \textbf{Score} \\ 
\midrule
M1 & Appropriateness of Proactive Guidance & Whether proactive questions or guidance match the user's context and needs & Human rating (1--5 scale) & 4.7 \\
M2 & Effectiveness of Explanations & Whether explanations help the user understand and reduce confusion & Human rating (1--5 scale) & 4.8 \\
M3 & Accuracy of Fact Verification & Whether dialogue content aligns with factual information & Human rating (1--5 scale) & 4.6 \\
M4 & Semantic Relevance & Semantic similarity between generated content and the user's context/input & Human rating (1--5 scale) & 4.7 \\
M5 & Discourse Coherence Score & Logical and semantic coherence of the dialogue &\textcolor{black}{ Coh-Metrix~\cite{graesser2004coh}} (-1--1 scale) & 0.92 \\
    \bottomrule
\end{tabular}
}
    \vspace{-10pt}
\end{table*}

\textcolor{black}{To assess the influence of model depth, we evaluated SFR-Net variants with one to seven layers under the same setting. As shown in Table~\ref{tab:depth_compare} and Figure~\ref{fig:roc_curves_compare}(d), performance is not monotonic with depth. The five-layer model achieves the best overall results, including an accuracy of \(0.804\pm0.016\), an F1 score of \(0.804\pm0.017\), and an AUC of 0.94.  Six or seven-layer variants bring no significant performance improvement and instead increase parameter count and inference time. Therefore, five layers provide the best trade-off between performance and computational complexity.}

\textcolor{black}{Beyond average performance, we further assessed model stability across individuals through a subject-specific variability analysis of 30 participants. As shown in Figure~\ref{fig:roc_curves_compare}(e), the mean subject-wise accuracy is 0.838 with a standard deviation of 0.081, indicating moderate inter-subject variability. The accuracy ranges from 0.667 to 0.958, suggesting that SFR-Net maintains generally stable performance for most participants despite individual differences.}

To promote reproducibility and further research, the source code and experimental data for the proposed SFR-Net have been made publicly available at our GitHub repository~\cite{SFRNetCode}.

\vspace{-10pt}

\subsection{Evaluation of ``4$W$-UT'' Prompt Controlled Proactive Chatting}

Given the variability and potential hallucinations of LLM outputs, we evaluated the stability and reliability of the conversational responses generated under the proposed ``4$W$-UT'' prompt. Using the beverage recognition scenario as a representative case, we generated 50 conversation samples with identical ``4$W$-UT'' prompts, and the generated samples were made available online for reference~\cite{pvcare_llm_agent}. Ten healthy participants reviewed all 50 samples, resulting in 500 human ratings for each subjective metric. As shown in Table~\ref{tab:chat_eval_metrics}, M1--M4 were subjective metrics scored on a 5-point scale, including the appropriateness of proactive guidance, effectiveness of explanations, factual accuracy, and semantic relevance. \textcolor{black}{Meantime, M5 was an objective discourse coherence score automatically computed using Coh-Metrix~\cite{graesser2004coh} to assess the logical and semantic coherence of the generated dialogue.} The results indicate that combining the LLM with the structured ``4$W$-UT'' prompting framework enables PV-Care to provide acceptable proactive assistance.

    \vspace{-10pt}

\subsection{Overall Performance Assessment for PV-Care Scheme}

\subsubsection{Hardware Implementation of the PV-Care Prototype and Participant Selection for Subjective Evaluation}

In addition to evaluating the performance of the proposed SFR-Net for EEG analysis, we further assessed the overall system feasibility of the PV-Care scheme through hardware implementation. \textcolor{black}{To validate the practicality of the proposed PV-Care, we developed a functional prototype focused on environmental sensing and EEG data acquisition, the implementation of the PV-Care prototype was detailed in Section II.A of the Supplemental Material.}

\begin{figure}[!t]
    \centering
    \includegraphics[width=0.9\linewidth]{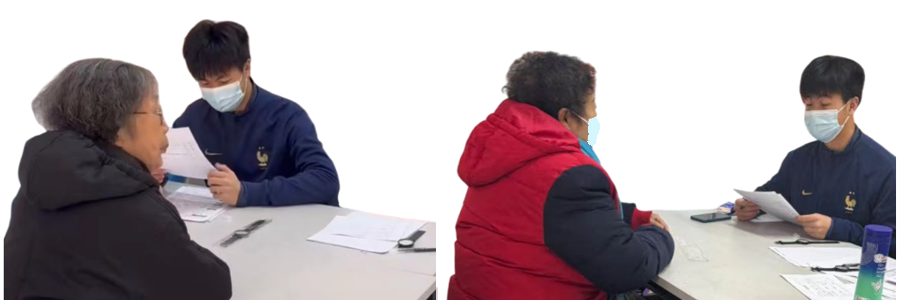}%
    \caption{MoCA and MMSE cognitive screening for participant selection.}
    \label{fig:screnn}
        \vspace{-10pt}
\end{figure}

Since PV-Care is designed primarily for individuals with MCI, we recruited suitable participants in collaboration with the Brain and Behavior Research Institute (BABRI) \cite{BABRI2021}, following established ethical guidelines. Cognitive screening was conducted using two standardized tools: the Montreal Cognitive Assessment (MoCA) and the Mini-Mental State Examination (MMSE). The MoCA is known for its high sensitivity in detecting MCI by assessing memory, language, attention, and executive functions \cite{nasreddine2005moca}, while the MMSE provides a general evaluation of cognitive status and is widely used in dementia screening \cite{folstein1975mmse}.

Sixty volunteers from local communities completed the MoCA and MMSE assessments (Figure~\ref{fig:screnn}). Table~\ref{tab:patient_scores} summarizes the cognitive and demographic profiles of the participants. The mean MMSE and MoCA scores were 26 and 23, respectively, confirming mild cognitive decline in the group. To ensure effective participation in the usability study, 20 individuals with MoCA scores between 18 and 25 were selected \textcolor{black}{for} our subjective evaluation of PV-Care. These participants demonstrated sufficient cognitive capacity to follow instructions and engage with the system.

\subsubsection{Prototype Testing and Usability Evaluation}

To evaluate the usability of PV-Care, we invited 20 participants selected through the screening process described above. \textcolor{black}{These participants had relatively preserved cognitive ability and were able to understand the experimental instructions and provide reliable subjective feedback. In a controlled environment, the selected participants were asked to wear the PV-Care prototype and complete two representative guided assistance tasks, corresponding to the \textit{Learning} and \textit{Memory Recall} cases, respectively. The detailed procedures of these two simulated cases are provided in Section II.B of the Supplemental Material.} After completing the simulated interactions, the participants filled out a subjective evaluation form (Table~\ref{table:evaluation_form}), which assessed wearing comfort, ease of use, prompt clarity, overall satisfaction, and willingness to use the system in daily life. The average scores were 4.7 for wearing comfort, 4.8 for ease of use, 4.6 for prompt clarity, 4.8 for overall satisfaction, and 4.9 for willingness to continue using the system.

\begin{table}[t!]
\centering
\caption{Subjective Evaluation of PV-Care Prototype (1 = low, 5 = high).}
\label{table:evaluation_form}
\footnotesize{
\setlength{\tabcolsep}{3pt}
\begin{tabular}{@{}llc@{}}
    \toprule
    \textbf{Metric} & \textbf{Remarks} & \textbf{Avg. Score} \\
    \midrule
    Wearing comfort    & Comfort while wearing the device & 4.7 \\
    Ease of use        & User-friendliness and interaction simplicity & 4.8 \\
    Prompt clarity     & Clarity of system prompts and feedback & 4.6 \\
    Overall rating     & General assessment of system performance & 4.8 \\
    Willingness to use & Interest in continued usage in daily life & 4.9 \\
    \bottomrule
    \end{tabular}
}
    \vspace{-10pt}
\end{table}

\subsection{Discussion and Limitations}

\textcolor{black}{Although the proposed PV-Care scheme demonstrates the feasibility of using wearable EEG for proactive MCI assistance, several limitations should be noted. 
Compared with high-density 64-channel EEG systems, the current low-density wearable setting provides limited spatial information and restricts the use of cross-channel analysis methods that rely on rich inter-channel relationships. 
Therefore, SFR-Net mainly focuses on temporal and frequency-domain features extracted from low-density EEG signals. 
Although previous studies report that consumer-grade dry-electrode EEG devices may have signal-quality limitations compared with gel-based EEG systems \cite{kleeva2024resting}, the Muse 4-channel wearable EEG device is selected in this work mainly because of its lightweight wearable form factor and suitability for daily-use scenarios. 
It should also be noted that Muse is not the only possible EEG sensing option for PV-Care. 
In future work, we will further evaluate other wearable EEG devices, such as the PSBD EEG Headband Pro \cite{PSBD}, and investigate lightweight cross-channel modeling strategies to improve brain-state recognition while maintaining wearable comfort and usability.}

\textcolor{black}{In addition, the current PV-Care prototype integrates cloud-based LLM services to provide context-aware voice assistance, but precise control over LLM-generated responses still requires further optimization. 
More refined prompt design, additional interaction data, and an improved agent workflow are needed to enhance the stability, safety, and personalization of the generated assistance. 
In the current work, we mainly verify the basic functionality and feasibility of the proposed prototype. 
More objective usability indicators, such as task completion rate, response time, error rate, and the number of required assistance prompts, will be further evaluated in future system studies.}

\textcolor{black}{For the application scope, our work mainly focuses on MCI users at the current stage, but the proposed proactive assistance framework may also be useful for other populations with cognitive decline or daily assistance needs, such as users with dementia. 
Since the cognitive conditions and assistance requirements of these users can vary significantly, future work will recruit more diverse participant groups and further adapt the EEG recognition, prompt generation, and interaction workflow to broaden the applicability of PV-Care.}
    \vspace{-5pt}

\section{Conclusion} \label{sec:conclusion}

\textcolor{black}{We proposed PV-Care, a proactive assistance scheme for individuals with MCI. 
PV-Care integrates wearable 4-channel EEG sensing, environmental perception, and large language model-based dialogue to provide real-time and personalized support. 
The 4-channel EEG setting is adopted for its convenience and practicality in daily use, while the proposed SFR-Net enables brain-state recognition under this low-density EEG condition. 
In addition, the visual sensing module integrated with the EEG device captures surrounding images and combines visual-content analysis with the recognized brain-activity state. 
The structured ``4$W$-UT'' prompts are designed based on the practical needs of MCI assistance and are used to set up the LLM to generate context-aware responses by combining the user state, environmental information, and personal context. 
Simulation experiments and user evaluations confirm the system's usability, EEG-based three-state brain-activity recognition performance, and conversational effectiveness. 
In future work, we will further improve EEG recognition robustness, collect more data from elderly and MCI participants, and optimize the prompt generation and agent workflow for more reliable daily support.}
    \vspace{-10pt}
\section*{Acknowledgment}
\textcolor{black}{The authors gratefully acknowledge the anonymous reviewers for their valuable comments and constructive suggestions. 
The authors also sincerely appreciate the opportunity to participate in the  research project jointly conducted by Beijing Normal University and Fudan University. 
The idea of PV-Care was inspired by the authors' experience in this cognitive research project, especially during the community-based volunteer screening activities. 
Sincere thanks are extended to Prof. Zhineng Chen from Fudan University for his guidance on this project, and to Prof. Xin Li from Beijing Normal University for her support in participant recruitment (ethically allowed) and related experimental conditions. The authors would also like to thank Prof. Yaguang Zhang from Purdue University for valuable discussions on this research and manuscript preparation.}
    \vspace{-5pt}

\bibliographystyle{IEEEtran}
\bibliography{bibtex/citations}

\newpage
\section{Supplementary Materials} \label{sec:supplementary}
\subsection{EEG Signal Acquisition and Experimental Dataset} \label{sec:eeg-acquisition}

\begin{figure}[t!]
\centering
\includegraphics[width=0.9\linewidth]{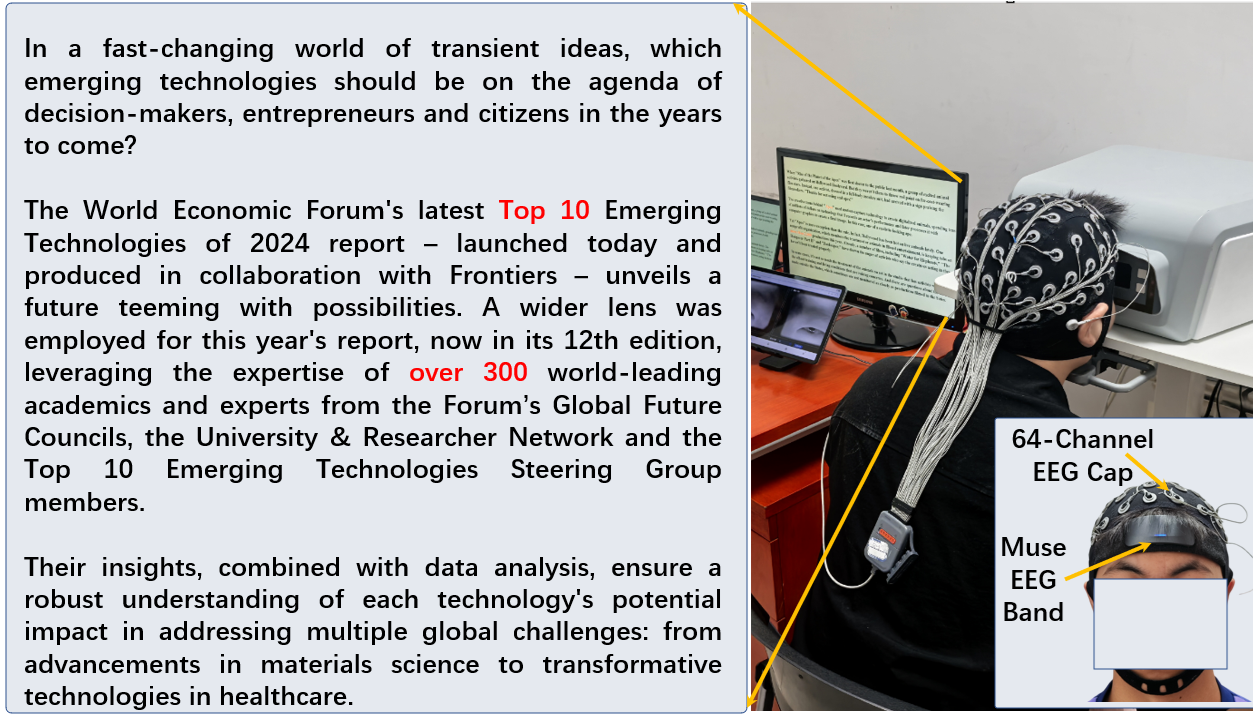}
    \vspace{-5pt}
\caption{Experiment for collecting EEG during learning activities.}
\label{fig:EEG-test}

    \end{figure}

More than 30 volunteers participated in our EEG sensing experiments to establish a comprehensive dataset. Each participant wore both the wearable 4-channel EEG device and the standard 64-channel EEG device throughout the entire data collection process, as illustrated in the lower-right part of Figure~\ref{fig:EEG-test}.

\begin{figure}[t!]
\centering
\includegraphics[width=0.9\linewidth]{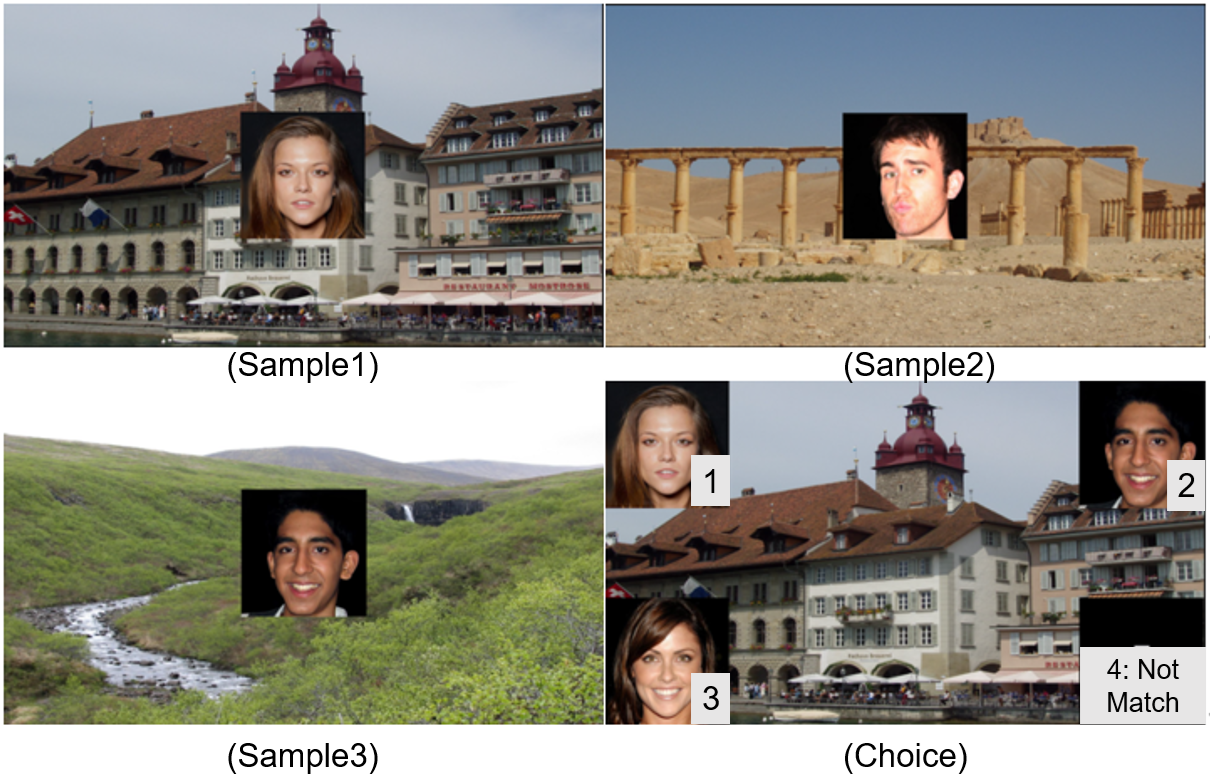}
    \vspace{-10pt}
\caption{Experiment for collecting EEG during memory recall activities (participants matched background images with memorized faces).}
\label{fig:mem-test}

\end{figure}

\textcolor{black}{The EEG data acquisition experiments were designed as follows:} 3-Minute Resting State Session $\rightarrow$ 5-Minute Learning Task $\rightarrow$ 3-Minute Resting State Session $\rightarrow$ 5-Minute Memory Recall Task $\rightarrow$ 3-Minute Resting. This experimental protocol ensures that the necessary data for all three states are collected in a single session, adhering to standard EEG data collection procedures. The specific tasks for each state are described as follows:

\begin{itemize}
    \item \textit{Resting State}: Participants were instructed to remain seated and relaxed with their eyes open for approximately 3 minutes to establish a baseline EEG signal. \textcolor{black}{These resting-state data serve}  as a reference for each subject.
    
    \item \textit{Learning State}: Participants were presented with a visual learning task. They watched a short news video with highlighted text (Figure~\ref{fig:EEG-test}), and participants were required to understand and learn the highlighted content. EEG data collected during this period were labeled as learning state. 
    
    \item \textit{Memory-Recall State}: After a 3-minute resting interval, participants were asked to memorize a series of face–background pairs. During the recall phase, background images were presented, and participants were asked to select the corresponding memorized face. For example, as shown in the lower-right “Choice” panel of Figure~\ref{fig:mem-test}, the correct selection is label 1, which matches the face–background pair in the upper-left “Sample 1” panel. When the participant correctly selected the face corresponding to the original sample image, the participant’s EEG data were stored as memory recall state.
\end{itemize}

\textcolor{black}{Although both the 64-channel EEG signals and the 4-channel Muse EEG signals were recorded simultaneously in our experiment, the target EEG input of PV-Care is the low-density 4-channel Muse data. Due to possible instability in participant performance and EEG data acquisition, we first used the simultaneously recorded 64-channel EEG signals to verify the validity of each data segment. Specifically, we examined whether the 64-channel EEG data showed distinguishable patterns among the three states: \textit{Resting}, \textit{Learning}, and \textit{Memory Recall}. Since the 4-channel Muse data were time-synchronized with the 64-channel recordings, the corresponding 4-channel segments were retained only when the same time periods were validated by the 64-channel EEG analysis. These valid 4-channel EEG segments were then used to construct the dataset for training and evaluating the proposed SFR-Net model.}


    \vspace{-5pt}
\subsection{PV-Care Prototype Hardware and Case Study} \label{sec:prototype}

\begin{figure}[t]
    \centering
 \includegraphics[width=0.9\linewidth]{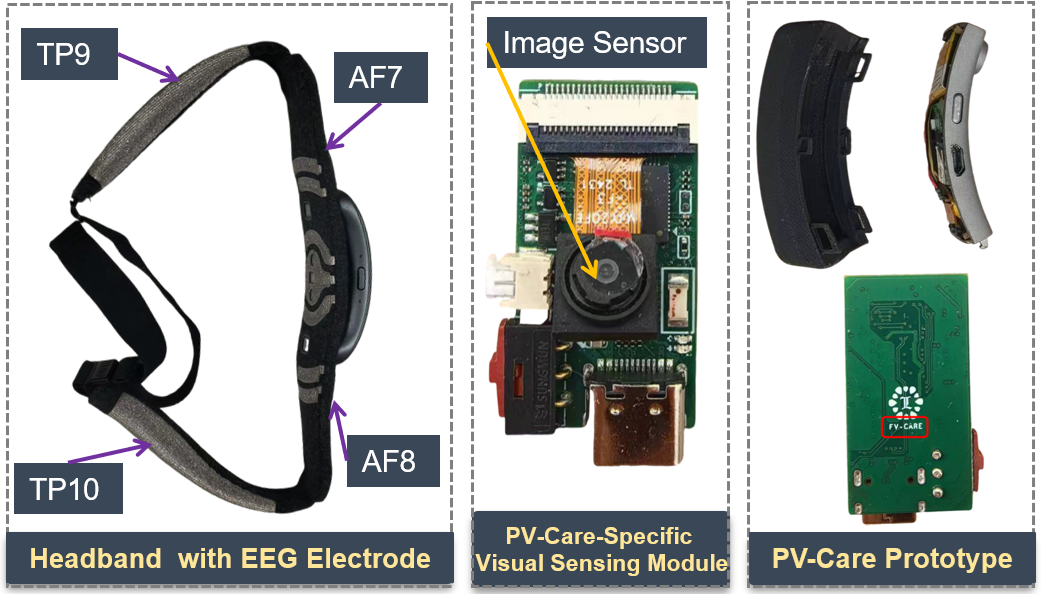}%
    \caption{Wearable PV-Care prototype.}
    \label{fig:equip}
    \vspace{-5pt}
\end{figure}

\subsubsection{Hardware Implementation of the PV-Care Prototype}

To validate the practicality of the proposed PV-Care scheme, we developed a functional prototype of it. We adopted a commercially available 4-channel wearable EEG headset (Muse), which is capable of capturing real-time brainwave signals. To meet the integrated and wearable design requirements of PV-Care, we specially developed a lightweight camera module capable of capturing 640$\times$480 resolution images for real-time environmental sensing. This visual sensing unit was mechanically integrated with the EEG headset via a custom-designed 3D-printed interface and assembly mechanism, ensuring stable alignment during operation while maintaining user comfort. Furthermore, by equipping the system with a compact rechargeable power supply, we constructed a fully functional prototype of the PV-Care sensing hardware, as illustrated in Figure~\ref{fig:equip}.

    \vspace{-5pt}
\subsection{Case Study: Assistant Chatting Powered by ``4$W$-UT'' Prompts}

To evaluate the functionality of the PV-Care chatting agent (In this experiment, we used ChatGPT APIs of OpenAI as the LLM to support our chatting agent.), we conducted two representative simulation scenarios that tested the system’s ability to detect users’ cognitive states, construct contextual prompts, and deliver proactive, context-aware assistance. Volunteers were invited to participate and provide feedback. 

\subsubsection{Scenario 1: Assisting with Beverage Recognition and Learning}

\begin{figure}[!t]
    \centering
    \includegraphics[width=1\linewidth]{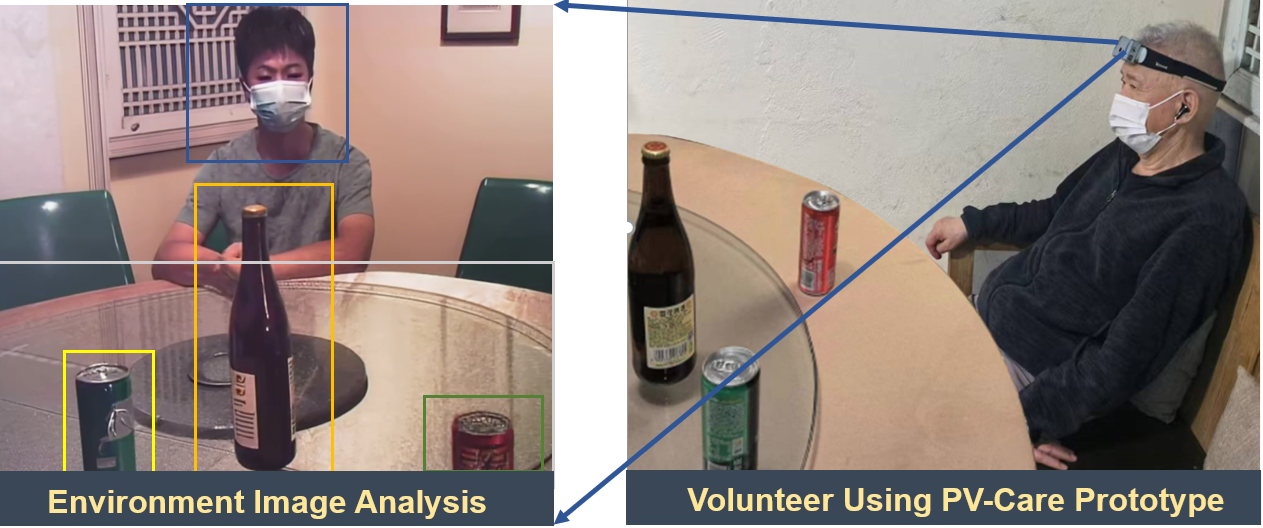}%
    \caption{(a) Environmental image captured by PV-Care. (b) Volunteer wearing the PV-Care prototype during testing.}    
    \label{fig:case}
        \vspace{-10pt}
\end{figure}

As shown in Figure~\ref{fig:case}, a volunteer wearing the PV-Care prototype was seated at a round table with three beverages: a bottle of beer, a can of Coke, and a can of Sprite. To simulate a realistic home scenario, another participant—a registered family member named Leonardo—was seated opposite the volunteer.

Before initiating any conversation, PV-Care performed a multi-stage perception process. The SFR-Net identified that the user was in a \textit{Learning} state based on EEG signals, indicating a focus on the objects in front. Meanwhile, the wearable camera captured the surrounding scene, and samartphone based visual analysis recognized Leonardo. The system then constructed a ``4$W$-UT'' contextual prompt, which was transmitted to ChatGPT using the \textit{system} role (see Section II.D for details). Based on the generated response, PV-Care proactively initiated a context-aware conversation, as shown below:
\begin{enumerate}
    \item \textbf{PV-Care:} \textit{I see you’re looking at the three drinks in front of you, and Leonardo is here with you. Would you like some help understanding what they are?}

    \item \textbf{PV-Care:} \textit{[Waiting for user response...]}

    \item \textbf{PV-Care} (if no response within 10 seconds, re-initiates the prompt): \textit{I can help you learn more about the drinks you're looking at—just let me know when you're ready.}

    \item \textbf{Loop:} repeat 2), 3) processes  until a response is received from the user.

    \item \textbf{PV-Care} (upon receiving user response, continues with contextual explanation): \textit{The can with the red label is Coke. It’s a sweet, caffeinated soft drink....}
    \item \textbf{User:} \textit{ I want....}
\end{enumerate}

This case demonstrates that PV-Care can recognize when the user is in a \textit{learning} state and provide targeted real-world assistance, enabling users to better understand and interact with their environment.

\subsubsection{Scenario 2: Reminding and Guiding the User to Return Home}

The second scenario simulated a safety-critical situation for an MCI user. A volunteer equipped with the PV-Care prototype walked outdoors and stopped at a crossroads, attempting to recall the correct direction home. At this moment, SFR-Net detected a \textit{Memory Recall}  state. Simultaneously, the wearable camera captured nearby visual landmarks, and the GPS module determined the user's exact location. Combining this information with stored home address data, PV-Care constructed a safe navigation route and generated a contextual 4$W$-UT prompt for the language model:

\begin{quote}
\textit{``role: system, content: I am on Livernois Road with GPS ddmm,N/S, dddmm,E/W. Current time is year-month-day-time. In front of me: road, tree, building. The user is currently in a memory recall state.''}
\end{quote}

PV-Care then initiated a proactive navigation dialogue:

\begin{enumerate}
    \item \textbf{PV-Care:} \textit{You seem to be near the park. Are you trying to find your way home?}
    
    \item \textbf{PV-Care:} \textit{[Waiting for user response...]}

    \item \textbf{PV-Care} (if no response within 10 seconds, re-initiates \textcolor{black}{the } prompt): \textit{If you're unsure which direction to go, I can guide you back home. Just let me know when you're ready.}

    \item \textbf{Loop:} repeat 2), 3) until a response is received from the user.

    \item \textbf{PV-Care} (upon receiving user response, proceeds with personalized navigation instructions): \textit{From your current location, turn right at the next intersection. Then continue straight for about 300 meters until you reach Timberview Street. Turn left there—your home is just around the corner.}
     \item \textbf{User:} \textit{ Which way...}
\end{enumerate}

This scenario demonstrates PV-Care’s capability to identify \textit{memory recall} states, contextualize the surrounding environment, and provide timely, back-home navigation.

\end{document}